\documentclass[10pt,nocopyrightspace]{sigplanconf}

\pdfoutput=1

\usepackage{listings}
\usepackage[hyphens]{url}
\usepackage[hidelinks]{hyperref}
\usepackage{upquote}
\usepackage{color,soul}
\usepackage{graphicx}
\usepackage{dcolumn}
\usepackage{amsmath}
\usepackage{amssymb}
\usepackage{textcomp}
\usepackage{stmaryrd}
\usepackage{wasysym}
\usepackage{cancel}
\usepackage{eufrak}
\usepackage{colortbl}
\usepackage{subfigure}
\usepackage{alltt}
\usepackage{lineno}
\usepackage{comment}

\usepackage{languages}

\newcommand{\sv}[1]{{\small\texttt{\tt #1}}}

\newcommand{\ucm}{jUCM}

\begin{document}

\setlength{\pdfpageheight}{\paperheight}
\setlength{\pdfpagewidth}{\paperwidth}

\conferenceinfo{CONF 'yy}{Month d--d, 20yy, City, ST, Country} 
\copyrightyear{20yy} 
\copyrightdata{978-1-nnnn-nnnn-n/yy/mm} 
\doi{nnnnnnn.nnnnnnn}

\title{\ucm{}: Universal Class Morphing (position paper)}

\authorinfo{Aggelos Biboudis}
  {University of Athens}
  {biboudis@di.uoa.gr}
\authorinfo{George Fourtounis}
  {University of Athens}
  {gfour@di.uoa.gr}
\authorinfo{Yannis Smaragdakis}
  {University of Athens}
  {yannis@smaragd.org}

\maketitle

\begin{abstract}
  We extend prior work on \emph{class-morphing} to provide a more
  expressive pattern-based compile-time reflection language. Our \emph{MorphJ}
  language offers a disciplined form of metaprogramming that produces types by statically
  iterating over and pattern-matching on fields and methods of other types. We
  expand such capabilities with ``universal morphing'', which also allows
  pattern-matching over types (e.g., all classes nested in another, all supertypes of
  a class) while maintaining modular type
  safety for our meta-programs. We present informal examples of the functionality
  and discuss a design for
  adding universal morphing to \java{}.
\end{abstract}

\section{Introduction}

The ultimate flexible software component is one that safely adapts its
behavior \emph{and} its interface depending on its uses. When a
component's interface is statically defined (as in the case of classes
in a statically typed language), such adaptation requires a
meta-programming facility. Meta-programming is typically low-level and
unwieldy, with few guarantees of safety. Mechanisms for compile-time
reflection~\cite{huang_morphing:_2011,ctr} have been proposed to
address such safety needs.

In our previous
work~\cite{huang_expressive_2008,huang_morphing:_2011,gerakios_forsaking_2013}
we presented and extended \emph{MorphJ}. MorphJ is a language that
adds compile-time reflection capabilities to \java{}. A programmer is
able to capture compile-time patterns and encode them in
(meta-)classes. 
Each pattern is associated with a generative scenario. For instance, a
morphed class \sv{Listify} may statically iterate over all the methods
of another, unknown, type, \sv{Subj}, pick those that have a single argument,
and offer isomorphic methods: whenever \sv{Subj} has a method with
argument \sv{A}, \sv{Listify} accepts a \sv{List<A>} . (The implementation
of every method in \sv{Listify} can then, e.g., iterate over all list
elements, and manipulate them using \sv{Subj}’s methods.) 

\begin{lstlisting}[style=java]
class Listify<Subj> {
  Subj ref;
  Listify(Subj s) {ref = s;}

  <R,A>[m] for (public R m(A): Subj.methods)
  public R m (List<A> a) {
    ... /* e.g., call m for all elements */
  }
}
\end{lstlisting}

MorphJ offers program transformation capabilities but with modular
type-safety guarantees: type-checking (via MorphJ) the code of
\sv{Listify} guarantees that all the classes it may produce (for any
type \sv{Subj}) also type-check (via the plain Java type system).

In this work we complement MorphJ with the ability to statically
reflect over classes, instead of just fields and methods. We discuss
our early motivation with examples over nested classes.

\section{Application: (Static) Nested Classes}

Classes are the typical unit of modularity in an object-oriented
language.  To form larger modules, one can group classes together into
components such as packages, or assemblies. At the language level, the
class mechanism itself can serve as a component, encapsulating other
classes. This is elegant from a modeling standpoint (a single concept
for all levels of modularity) and even captures existing language
features that allow the nesting of classes.

Nested classes can be either inner classes or static nested classes in
\java{}. Folklore in the \java{} community suggests to favor static
nested classes over inner classes and use the latter only if it is
absolutely needed (Item 22 in~\cite{bloch_effective_2008}).
Programmers use static nested classes in various practical
scenarios. In compiler engineering, static nested classes are usually
used when representing abstract syntax tree (AST) nodes. \sv{javac} in
fact, contains static nested classes for AST nodes that also extend
the top-level class,
\sv{JCTree}.\footnote{\url{http://hg.openjdk.java.net/jdk8/jdk8/langtools/file/jdk8-b132/src/share/classes/com/sun/tools/javac/tree/JCTree.java}}
Tools such as ANTLR that generate parsers also generate code of this
form. In UI engineering, several tools generate class definitions that
contain static nested classes---e.g., the Android Asset Packaging Tool
that generates the \sv{R} class, a strongly-typed view of resource IDs
for all the resources in the resources
directory.\footnote{\url{http://developer.android.com/guide/topics/resources/accessing-resources.html}}

Our universal morphing techniques find interesting applications
in (static) nested classes.

\paragraph{Ex1. Replace inheritance with delegation for all classes in a
  library. } In this example we want to replace inheritance with delegation
automatically for all static nested classes of \sv{Library}. This feature is
offered as a refactoring mechanism in IDEs today but the user may need to
generate a delegation-view via an existing hierarchy for all classes. Such
existing hierarchy is enclosed in the class \sv{Library} below:

\begin{lstlisting}[style=java]
class Library {
  static class Vector {
    boolean isEmpty() {}
  }
  static class Stack extends Vector { }
}
\end{lstlisting}

The programmer's intention is to have a view of the library that relies on
delegation like the one below:
\begin{lstlisting}[style=java]
class Library {
  static class Vector {
    boolean isEmpty() {}
  }
  static class Stack {
    Vector subobject;
    boolean isEmpty() { subobject.isEmpty(); }
  }
}
\end{lstlisting}
We introduce the static~\sv{for} keyword for static reflection over classes. In line 2
of the \sv{LibraryDelegated} we use it to iterate over all classes in the type
\sv{Library}. The pattern that we look for is that of classes that extend some
other class. All classes inside \sv{L} that are going to be captured will have a corresponding
definition in \sv{Delegate<L>}. Inside each class definition we define a
\sv{subobject} field of type \sv{S} (the supertype). In lines 5-6 we rely on the
\sv{static-for} we introduced in MorphJ.
\begin{lstlisting}[style=java]
class Delegate<L> {
  <C,S> for (C extends S : L.classes) 
  static class C {
    S subobject#S;
    <R, A*> [m] for(public R m(A) : S.methods)
    R m(A a){ return subobject#S.m(a); }
  }
}
\end{lstlisting}

\paragraph{Ex2. Introduce interface and add a new method. } In the following we
introduce an interface that is implemented by all static nested classes. Again
this is realized by reflecting over all classes of the type that is going to
parameterize the \sv{AlertingGraph} type.

\begin{lstlisting}[style=Java]
interface Alert { void alert(); }

class AlertingGraph <class X> {
  [N] for (N : X.classes)
  static class N extends X.N implements Alert {
    [m] for(public void m () : N.methods)
    public void m() {
      alert();
      m();
    }
    void alert() { System.out.println("Alerted!"); }
  }
}
\end{lstlisting}

\paragraph{Ex3. Merge two classes into one (including nested classes).} 
We can create a highly generic class that consists of the union
of members (methods and classes) of two others, with one of them
taking precedence.

\begin{lstlisting}[style=Java]
class Union<class B, class C> {
  <R, A*> [m] for (R m(A) : B.methods)
  R m(A a) { super.m(a); }
  <R, A*> [m] for (R m(A) : C.methods; 
                   no R m(A): B.methods)
  R m(A a) { super.m(a); }

  [N] for (N : B.classes)
  class N {
    <R,A> [m] for (R m(A) : N.methods)
    R m(A a) { b.m(a); }
    <NB> for (NB : N.classes)
    class NB extends N.NB {  }
  }
  [N] for (N : C.classes; not N : B.classes)
  class N {
    <R,A*>[m] for (R m(A) : N.methods)
    R m(A a) { b.m(a); }
    <NC> for (NC : N.classes)
    class NC extends N.NC {  }
  }
}
\end{lstlisting}

There is a wealth of other examples of universal morphing. For
instance, we can iterate over all interfaces implemented by a class,
we can offer highly-generic \emph{mixin layers} \cite{mixinlayers},
we can scrap the traversal boilerplate in external visitor patterns.

\section{Conclusion} We are working on \ucm{}, an extension
of MorphJ that enables more compile-time reflection patterns. A major
challenge includes designing the type system extension that will
ensure modular type-safety of meta-programs.

\renewcommand{\baselinestretch}{0.9}
\renewcommand{\bibsep}{0.1em}
\renewcommand{\bibfont}{\footnotesize}

\subparagraph*{Acknowledgments.}  
We gratefully acknowledge funding by the Greek Secretariat for
Research and Technology under the ``MorphPL'' Excellence (Aristeia)
award.

\bibliographystyle{abbrvnat}
\bibliography{universal}

\end{document}